# A 32 Channel Time-Tagging and Coincidence Detector Unit with High Data Throughput

Dr. Attila Hidvégi

*Abstract*— Time-tagging units and coincidence detectors are used in many scientific research fields. The required timing resolution and number of input channels are varying, but some emerging experiments in the field of quantum optics require up to 32 input channels with a timing resolution of approximately 10 ps and high data processing capability. This work is about a custom designed FPGA based time-tagging and coincidence detector unit, with 32 input channels, 8 ps of timing resolution, high data processing capability and with high bandwidth communication ports, such as USB-3 and PCIe x4, for readout. With very high timing resolution and many channels it is crucial to properly characterize the performance of the time-to-digital converters in every input channel, to validate their accuracy. Important sources of error are discussed and a common method of performance measurement is evaluated, together with its often overlooked flaws. The performance measurement implemented in this project characterizes every channel simultaneously, with no additional external instrument required except for an extra crystal oscillator on the PCB and a built in pattern generator in the FPGA. The advanced measurement is accurate enough to detect different type of jitters, from varying sources, measure noise caused by power supplies, measure linearity and characterize every input channels in detail. The measurement results are presented and evaluated in detail. This board with its high performance, large number of input channels and detailed characterization is currently unique and cutting edge.

## I. INTRODUCTION

TIME-TAGGING and coincidence detector units are key instruments used in many scientific fields. Some examples are life science, material science, metrology, nuclear science and quantum optics, just to mention a few. The experiment that initiated this project is in the field of quantum optics, that needed many input channels to receive data from all the detectors simultaneously, but the board is generic and could be used for other scientific experiments as well.

The requirement for the system was to have 32 input channels, functions as both time-tagging unit and coincidence detector, have a timing resolution of approximately 10 ps, capability to processing a large number pulses, have 8 programmable pattern trigger outputs and have a high bandwidth readout port, such as USB-3.0 or PCIe. From the coincidence vectors histograms with programmable duration should be generated in real-time. Result data must be stored in real-time to some type of storage device. Most importantly the device needs to be properly characterized and the performance of each channel must be measured individually.

## II. BOARD OVERVIEW AND FEATURES

The main board consist of a custom designed PCIe form factor card, Fig. 1, based around the Artix-7 200 FPGA [1] from Xilinx. The FPGA receives input signals through 32 SMA connectors, where all traces are matched to equal length. All pulses are labeled with time-tags, of 8 ps time resolution, and processed by the coincidence processor in the FPGA. The results are transmitted over a USB-3.0 SuperSpeed (SS) connection, or a PCIe x4 Gen. 2 port, to a host computer for further data processing. The generated coincidence vectors are also compared to 8 programmable registers in the FPGA to generate external pattern trigger outputs through SMA connectors.

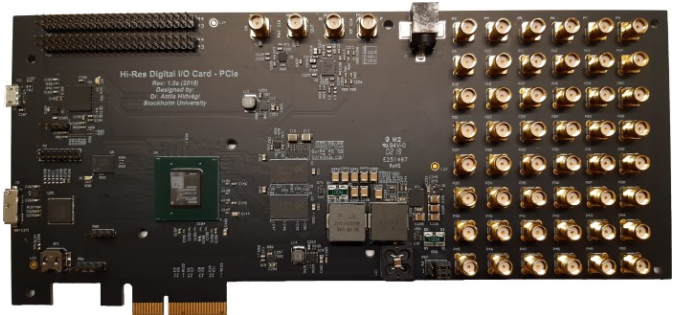

Fig. 1. *Custom PCIe board with SMA inputs and trigger outputs to the right, reference clock input/output on the top and USB-3 to the left.*

A high-performance fractional PLL frequency synthesizer [2] is implemented on the board, Fig. 2, to generate the high frequency RF clock for signal sampling in the FPGA. The PLL can synchronize to either a local high quality crystal oscillator or an external reference clock. The selected reference clock source is duplicated through a clock buffer, which simplifies synchronization across multiple board.

The board also contains 1 GB of DDR3L RAM, for local data storage and processing, plus an expansion port for future applications.

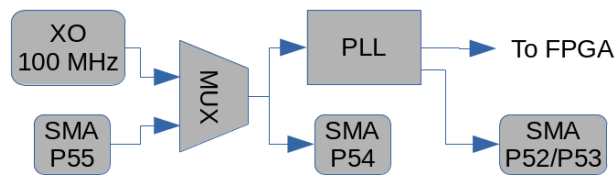

Fig. 2. Sampling clock scheme. The reference is either a local crystal oscillator or an external clock source.

### III. FPGA Architecture

The FPGA design consist of several subsystems, shown in Fig. 3. The first subsystem is a multi-channel time-tagging unit (MCTTU) that for each individual channel contains a time-to-digital converter (TDC), an individually programmable offset, an optional dead-time filter with programmable length and a FIFO buffer for temporary time-tag storage. The MCTTU transmits a time sorted stream of time-tags, bundled with its channel number, to the next subsystem. The subsystem that follows the MCTTU is the coincidence processor (CP). The CP processes the stream of time-tags and combines them into coincidence vectors. The length of the coincidence window is programmable in steps of 8 ps, with a maximum duration of 2.5 µs. An event filter discards all vectors that has less bits set than required by a programmable threshold. The remaining coincidence vectors are transmitted to the pattern trigger subsystem and the readout subsystem. The pattern trigger compares each coincidence vector to 8 programmable registers and generates external trigger outputs. The readout subsystem consist of a protocol generator and the communication interface. It receives the data stream from either the MCTTU or CP, depending on the requested run mode by the host computer. The protocol generator formats the output data stream and injects necessary periodic status data, which helps to minimize the load on the host computer during data processing. The FPGA system is capable of processing 200 million pulses per second on average, and capture shorter bursts with even higher rate.

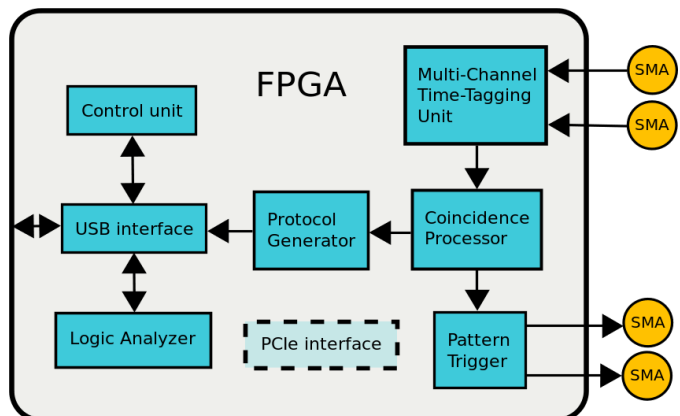

Fig. 3. Subsystems of the FPGA design.

### IV. Software Overview

The low-level software is a highly optimized multi-threaded program, which continuously receives data from the FPGA decompress it and process the data in real-time, and finally stores the results to a file or passing it to another receiving software. The board can run in different run modes: time-tagging mode, coincidence mode without time-tags, coincidence mode with time-tags and two real-time histogram generation modes. One of the histogram mode is using the accurate time-tags from the FPGA, while the other histogram mode is using the less accurate CPU time for its timing but with a higher data throughput from the FPGA. The maximum data rate from the board through the USB-3 port is about 80 million time-tag or coincidence vectors per second, but the PCIe port is capable of handling the full 200 million vectors per second data rate.

### V. Time-To-Digital Conversion

The TDC in the MCTTU is capable of capturing rising or falling edges of an input signal with a timing resolution of approximately 8 ps. It is based on the dedicated arithmetic carry chain [3] in the FPGA, which has a fast and dedicated signal path, utilized as delay elements. The propagation delay through each individual delay element is not uniform, however, causing nonlinearity problems. The solution for this design is the utilization of eight independent delay chains per input channel, which have each individual and unique nonlinearity characteristic, but when combined and averaged the resulting linearity is greatly enhanced. The delay elements, with 256 effective taps, are sampled with a frequency of 440 MHz. The sampling clock gives the course timing resolution and the signal propagation length through the delay chain gives the fine timing resolution.

### VI. Performance Measurements

Validation of the TDC performance is essential. One standard approach is to duplicate a source signal, from a pulse generator, with passive cabling and let two TDCs in the FPGA capture that. Since the variation of the propagation delay caused by thermal drifts through a short cable is insignificant, the relative offset between the two TDCs should remain constant. The observed jitter between the two TDCs should then represent the accuracy of the TDCs. This simplistic measurement is insufficient, however.

#### A. Contributions to the Error

There are different kind of errors contributing to the total jitter of a TDC. Some of them are completely invisible to the simplest measurement. The nonlinearity of an individual TDC is unique and therefore will directly contribute to the total jitter with a random component, which is always visible. The performance of the sampling clock, which has both random and deterministic components of jitter on its own, will directly affect all TDCs equally with systematic error and is completely invisible to the simplest type of measurement. Another important source of error contribution is caused by the signal path between the input pin of the FPGA and the delay elements, where the signal is captured. The longer path the signal has to propagate the more noise and random jitter it will accumulate inside the FPGA die. This factor, however, is most likely apparent on designs where some of the signals are inevitably long. With 32 input channels that is indeed the case, since most TDCs cannot be located directly nearby their associated input pins. This error contribution is clearly visible but requires all TDCs to be measured individually. Finally the noise on any power supply will impact the signal propagation speed inside the FPGA die, both on the signal path from the input pin to the delay elements and inside the delay elements themselves. Higher voltage will make the signal propagate faster and the characteristic noise from a DC/DC converter, supplying any part of the FPGA such as the internal core voltage and I/O voltage, will directly cause a deterministic jitter contribution to the signal, which will be an invisible systematic error to the simplistic measurement. However, varying signal length between the input pins and TDCs, between different input channels, can reveal the presence of power noise if measured correctly and accurately enough.

### B. Implemented Performance Measurements

The implemented performance measurements for this system tries to measure and characterize all the above mentioned error contributions. The large number of input channels, and the fact that every new FPGA firmware needs to be qualified, demands an automated analysis procedure.

To provide a test signal to all input channels simultaneously, during performance measurements, a pulse generator is implemented in the FPGA which provides the same test pattern to all the input pins. The pins are configured in bi-directional mode to allow the signal to pass through the I/O block, which guarantees that the whole input path is included in the measurement. The test pattern is a 2 MHz clock signal, provided by an external additional independent crystal oscillator, distributed through a global clock buffer within the FPGA for lowest possible clock skew and jitter contribution. Any jitter contribution from the global clock buffer will only add to the total jitter observed by the TDC. Every rising edge of the clock pattern will generate a time-tag from every inputs, with approximately constant offset between the channels. The average value of the time-tags, belonging to the same pulse, gives the best timing estimate of the rising edge. Comparing the offset of every channel to the best timing estimate of the rising edge reveals the jitter characteristic of each individual channel. By recording a large number of time-tags the time interval error (TIE) can be calculated and its frequency content can be analyzed for each channel.

The progression of the TIE can reveal potential drifts or deterministic jitter between the sampling clock and the independent clock, used to generate the test pattern. It does not reveal which of the two clocks contribute to the jitter, however, only that an independent measurement of the two clocks might be necessary.

Varying distances between the input pins and the TDCs on a 32 channel design magnifies the deterministic jitter caused by power noise. Frequency analysis of the TIE may pinpoint the source to a potential noise source.

The linearity of each channel is characterized by analysis of the statistical distribution of time-tags along the delay elements in the TDC, while the test pattern is sweeping through the delay chain.

All measurements and analysis are automated by the software, and a summary report is generated.

### VII. RESULTS

The first parameter and perhaps the most important one is the TDC linearity. Without a reasonable linearity all other measurements, which are based on the analysis of the time-tags, will be degraded and invalid. The measurements confirm that the linearity of the 256 taps in all TDCs are acceptable, and the averaging of eight individual delay chains within each TDC yields good accuracy. Fig. 4 shows the linearity of channel 19 as an example.

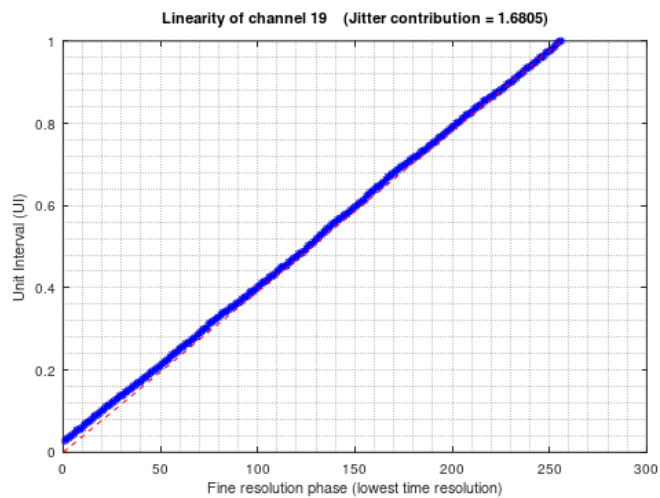

*Fig. 4. Linearity of TDC in channel 19. The x-axis is the tap number of the TDC while the y-axis is its equivalent position in time, expressed in unit-intervals.*

The second parameter of interest is the total RMS jitter of each individual channel, which indicates the true accuracy of the time-to-digital conversion. The measurements indicate an RMS jitter range of 1.9 - 4.5 LSB, corresponding to 16 - 36 ps, for different input channels. The theoretical minimal RMS error is approximately 0.5 LSB, corresponding to 4 ps, if no jitter of any kind or noise other than the quantization noise would be present. Fig. 5 shows the total RMS jitter of channel 19 as an example.

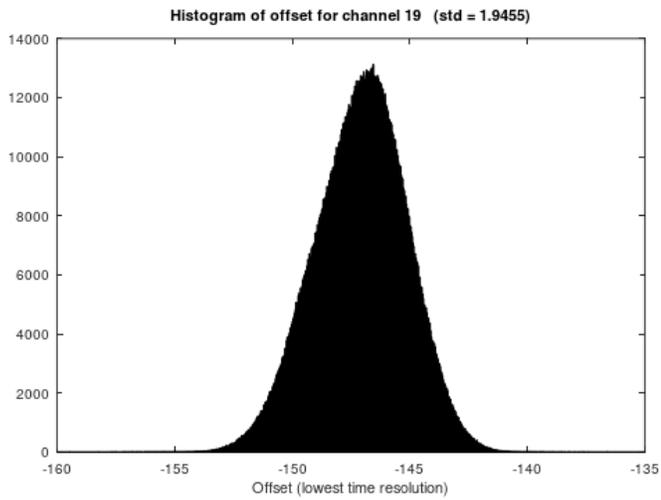

Fig. 5. Histogram of jitter in channel 19 with a total RMS jitter of 1.95 LSB, approximately equals to 16 ps.

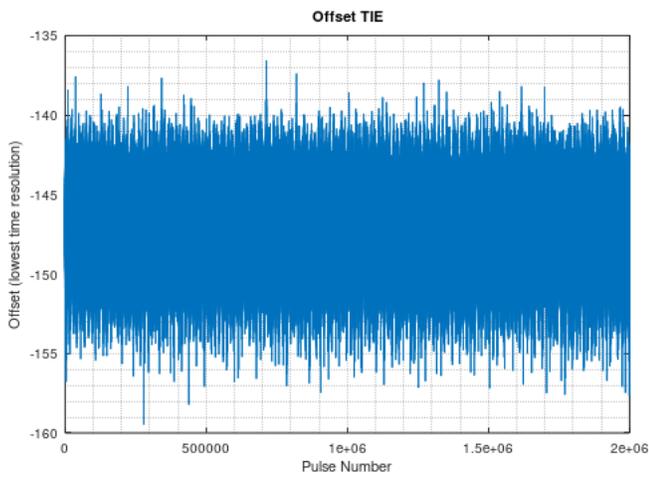

Fig. 6. Time Interval Error of channel 19, during 1 second. A stable offset during time progression indicates a stable sampling clock without drifts.

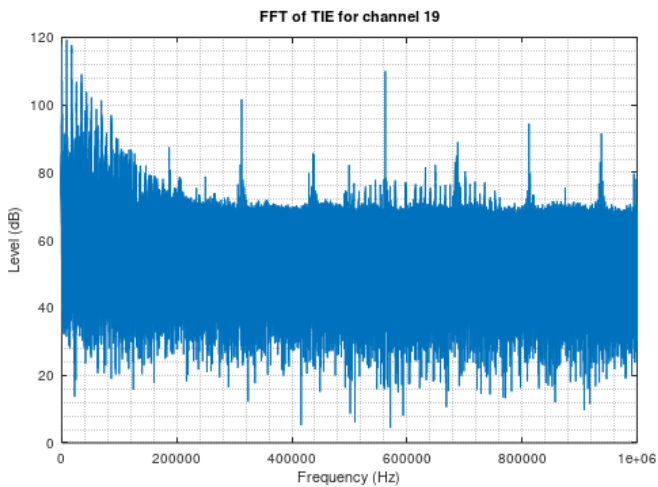

Fig. 7. Fourier transform of TIE for channel 19, corresponding 1 second of data capture giving 1 Hz frequency resolution.

Further analysis of the TIE diagram, Fig. 6, and its frequency components, Fig. 7, can separate the total RMS jitter into deterministic jitter components and a random jitter component. Two deterministic jitter components seems to dominate the total jitter. One at approximately 150 Hz, in Fig. 8, and another one at approximately 10 kHz, in Fig. 9. The random jitter component, in Fig. 10, is significantly less than its deterministic counterparts.

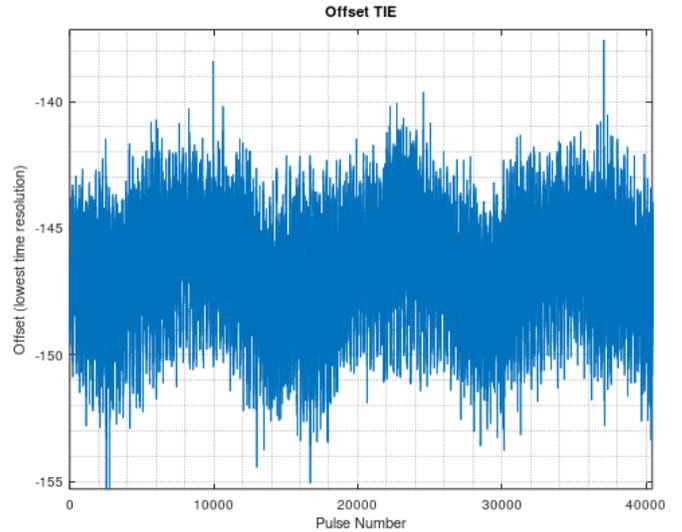

Fig. 8. Deterministic jitter with approximately 150 Hz and a peak-to-peak value of approximately 3 LSB, equal to 24 ps, on input channel 19.

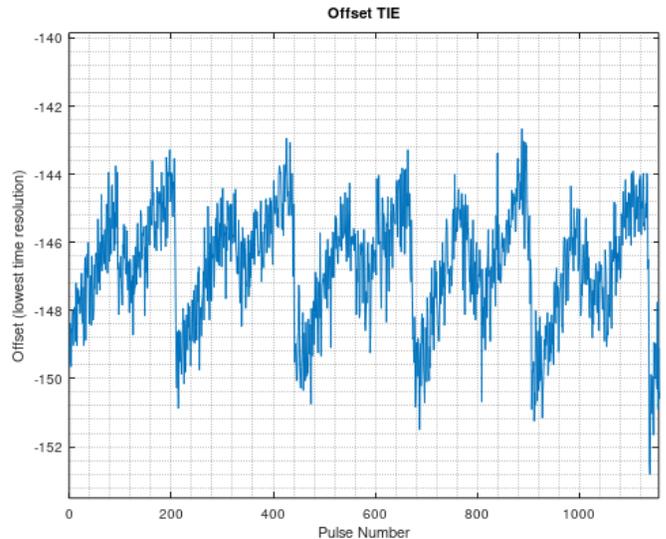

Fig. 9. Deterministic jitter with approximately 10 kHz and a peak-to-peak value of approximately 5 LSB, equal to 40 ps, on input channel 19.

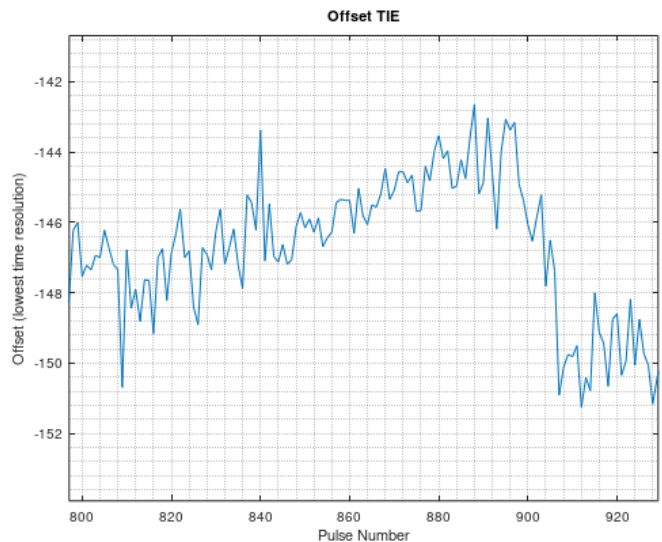

*Fig. 10. Random jitter on input channel 19, with significantly less amplitude than its deterministic counterpart.*

From these results it is possible to conclude that the TDC architecture is sufficiently accurate, with good linearity and low random jitter. The remaining deterministic jitter components have most likely power related causes. It is clear though that the noise does not originate from the DC/DC converter for the internal core voltage of the FPGA, which has a switching frequency of 500 kHz. The custom designed PCB has an extra noise filter for the internal core voltage capable of suppressing the noise by more than 60 dB at that frequency range. The power to the I/O banks of the FPGA does not have an additional filtering though, which could be the source to the deterministic jitter components. No early conclusions should be made, however, without further investigation. The source to the noise can be properly eliminated once identified.

## VIII. Summary and Conclusions

A time-tagging and coincidence detector unit, with 8 ps timing resolution, has been developed with a custom PCB design. It has 32 input channels and USB-3.0 or PCIe port for data readout. The large number of input channels, very high timing resolution, high data processing and high bandwidth readout capability makes this unit currently unique and cutting edge. The need for proper measurements and potential sources of problems have been discussed. Detailed performance measurements have been made with strong emphasis on correctness and completeness. The measurement results indicate very good performance, but there is room for further improvements. The very low random jitter, and a few minor deterministic jitter that should be possible to eliminate, should be a good reference for future designs. The board is generic and suitable for research in many scientific fields.